\documentclass[12pt]{article}
\usepackage{epsfig}

\def\beq{\begin{equation}}
\def\eeq#1{\label{#1}\end{equation}}
\def\eeqn{\end{equation}}

\def\beqa{\begin{eqnarray}}
\def\eeqa#1{\label{#1}\end{eqnarray}}
\def\eeqan{\end{eqnarray}}

\let\bar=\overbar

\def\Dslash{\not{\hbox{\kern-4pt $D$}}}
\def\dslash{\not{\hbox{\kern-2pt $\del$}}}

\def\msb{{\bar{\ssstyle M \kern -1pt S}}}

\def\Title#1{\begin{center} {\Large {\bf #1} } \end{center}}

\begin{document}

\Title{SUSY confronts LHC data}

\bigskip\bigskip


\begin{raggedright}  

{\it Jos\'e Francisco Zurita \index{Zurita, J.}\footnote{work done in collaboration with Marcela Carena and Eduardo Pont\'on.} \\
Institut f\"ur Theoretische Physik\\
Universit\"at Z\"urich \\
CH-8057 Z\"urich, SWITZERLAND}
\bigskip\bigskip
\end{raggedright}

\section{SUSY at the LHC}

There are many reasons why to expect Supersymmetry (SUSY) at the TeV scale: it solves the hierarchy problem, it provides a natural dark matter candidate, allows for gauge coupling unification, etc. Up to now the direct searches for SUSY, mainly based on missing transverse energy signatures have shown no significant excess over the SM backgrounds~\cite{susyexp}. 
While the actual bounds depend on many details, it is fair to say that the gluinos and the squarks of the first and second generation have to be heavier than 1 TeV, while the third generation squarks will be heavier than 200-300 GeV. The stop searches (For recent strategies 
see~\cite{stops}) will play a crucial role in determining if SUSY is indeed a natural solution to the hierarchy problem\footnote{After this talk was given, the LHC collaborations have presented results of their stop searches~\cite{stopsexp}, that do not alter the conclusions of the present discussion.} . 

Actually, the only requirements from naturalness~\cite{Lodone:2012kp} are a light stop ($m_{\tilde{t}} \le 400-500$ GeV) and light $\mu$ ($\le 200-300$ GeV). Such an scenario, dubbed as "Natural SUSY"~\cite{NaturalSUSY} is still an open possibility. 
This suggests the existence of two scales, $m_{\tilde{g},\tilde{q}} \geq  $ 1~TeV, and $m_{\tilde{t},\tilde{b}, \tilde{\chi} } \sim$ 200-400 GeV (see Ref.~\cite{Randall:2012dm} for single-scale Natural SUSY).

The Higgs boson provides an indirect way of searching for SUSY, and the recent discovery of a 125 GeV Higgs~\cite{LHCHiggs} with an enhanced diphoton rate, constrain even more the parameter space. We recall that in the Minimal Supersymmetric Standard Model (MSSM) at tree level the lightest Higgs mass fulfills $m_{h}^{0} \le m_Z | \cos (2 \beta) |$. The upper bound is zero for low $\tan \beta~(\approx 1)$, but it saturates for $\tan \beta > 10$.  In order to avoid LEP constraints~\cite{Schael:2006cr}, large radiative corrections to the Higgs mass are needed. These corrections depend strongly on several supersymmetric parameters, particularly on the stop mass and mixing angle, and prefer large $\tan \beta$ and $m_A > 300$ GeV .  The upper bound on $m_h$ in the MSSM is about 135 GeV~\cite{hmass}.

Much theoretical work has been focused in a 125 GeV Higgs in supersymmetric theories (see e.g ~\cite{125susyhiggs}). In brief, it is possible to obtain such a Higgs mass in the MSSM, but this only happens for specific values of the soft parameters (e.g large stop mixing, large $\tan \beta$, large $m_A$). The other important question is whether such a 125 GeV Higgs boson could have the observed rates to $\gamma \gamma$ and $ZZ$. This can be achieved, for instance, by having a light stau in the spectrum~\cite{staus}.

However, the Higgs can be pointing us toward more generic SUSY scenarios. On one hand the large $\tan \beta$ region is excluded by $\tau^+ \tau^-$ searches and, on the other hand low $\tan \beta$ is excluded due to the upper bound on $m_h$. This does not need to be the case in general extensions of the MSSM, and thus the Higgs sector provides information about beyond the MSSM (BMSSM) dynamics.

\section{Going Beyond: the BMSSM}

Extending the MSSM Higgs sector with dimension 5 operators, one finds~\cite{dim5}
\begin{equation}
\label{eq:dim5}
W=\mu H_u H_d + \frac{\omega_1}{2M} (1 + \alpha_1 X) (H_u H_d)^2 \, ,
\end{equation}
where $\alpha_1$ and $\omega_1$ are dimensionless, order one parameters,$X = m_s \theta^2$ is the so called "spurion superfield", that parameterizes SUSY breaking. In this talk we take $\mu=m_s = 200$ GeV and $M=1$ TeV, and we choose  $A_t=0$ (no mixing in the stop sector). The consequences of Eq.~(\ref{eq:dim5}) for the Higgs potential have been studied in~\cite{hdo}. 

At order $1/M^2$, one has many more new operators \cite{Carena:2009gx,Antoniadis:2009rn}. The collider phenomenology of dimension-six operators was studied in detail in Refs.~\cite{CPZ,Boudjema:2011aa}. Here we will show a few selected results.

\begin{figure}[!htb]
\begin{center}
\epsfig{file=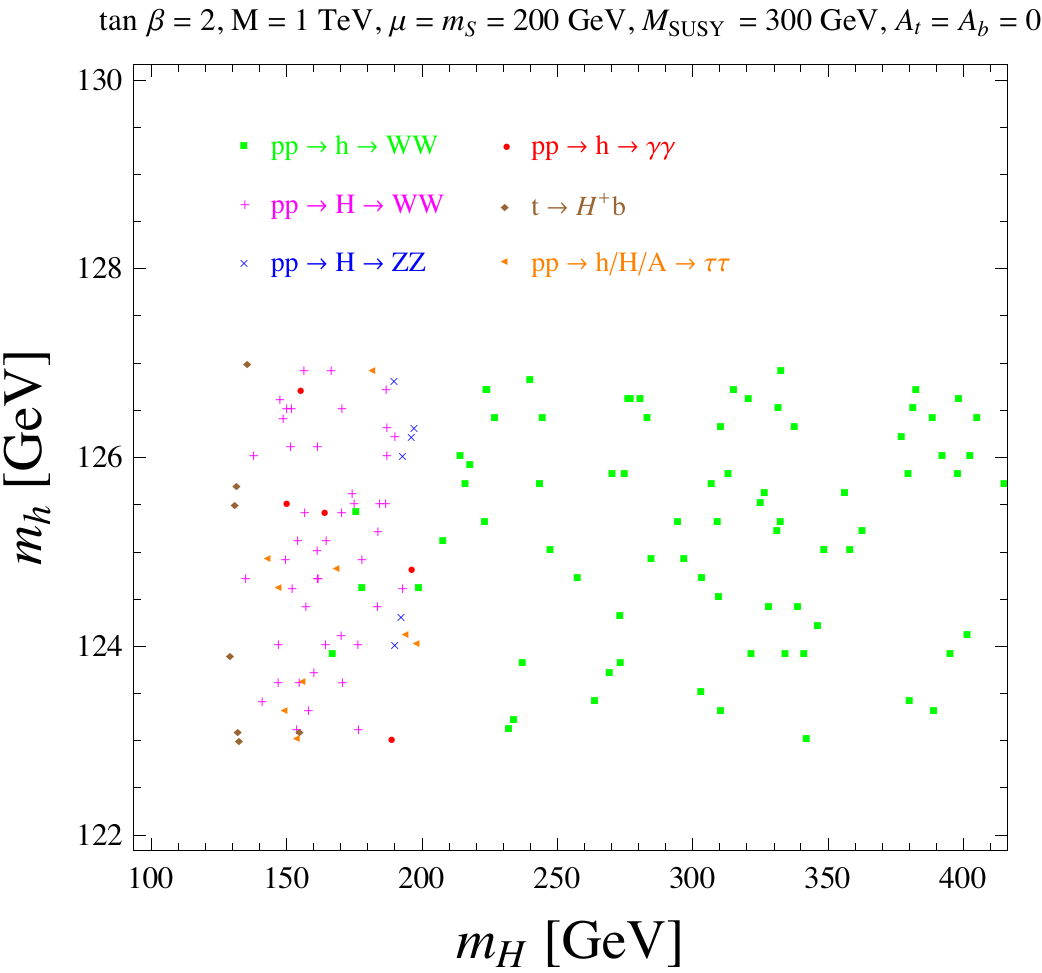,height=2.6in}
\epsfig{file=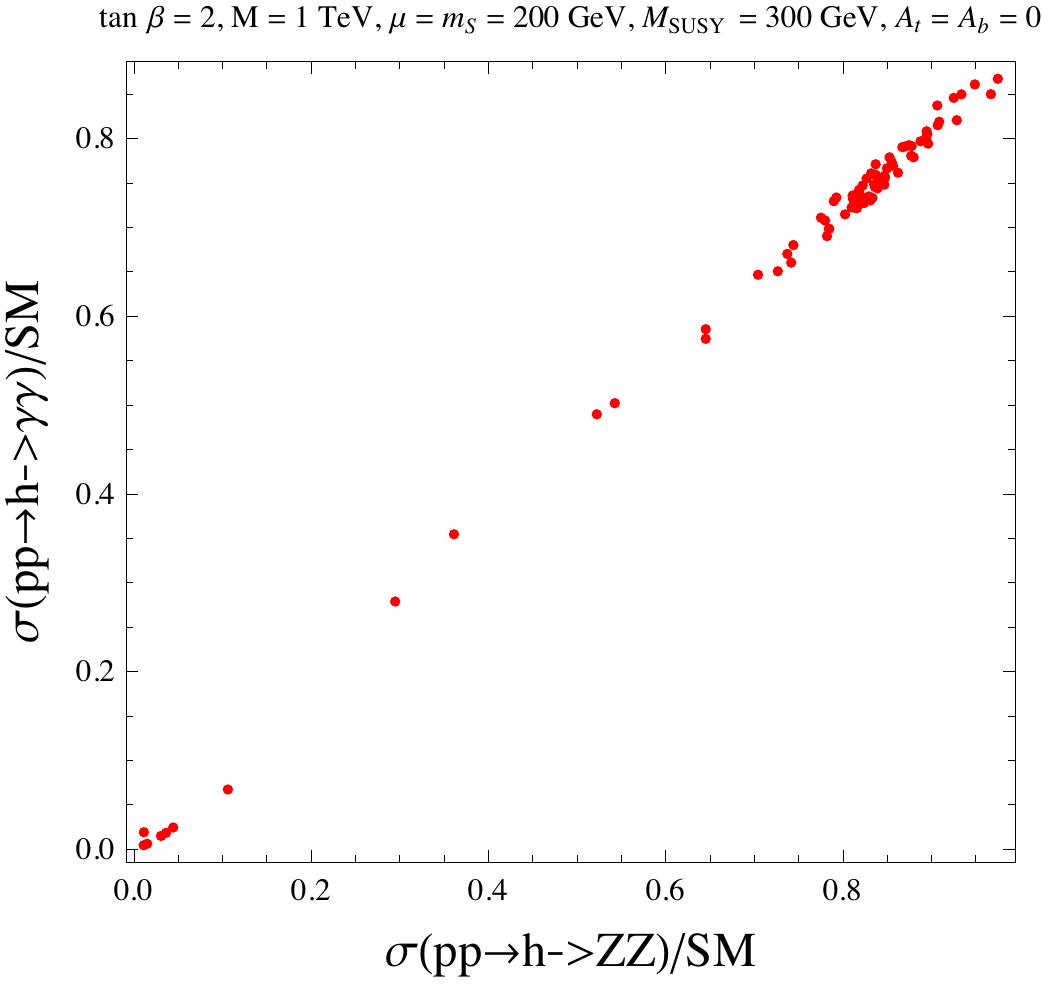,height=2.6in}
\caption{\small{(a) Experimentally allowed points in the $m_h-m_H$ plane, requiring $m_h \in$ [123-127] GeV and (b) Rate to $\gamma \gamma$ vs Rate to $ZZ$ for $\tan \beta=2.$}}
\label{fig:tb2}
\end{center}
\end{figure}
\begin{figure}[!htb]
\begin{center}
\epsfig{file=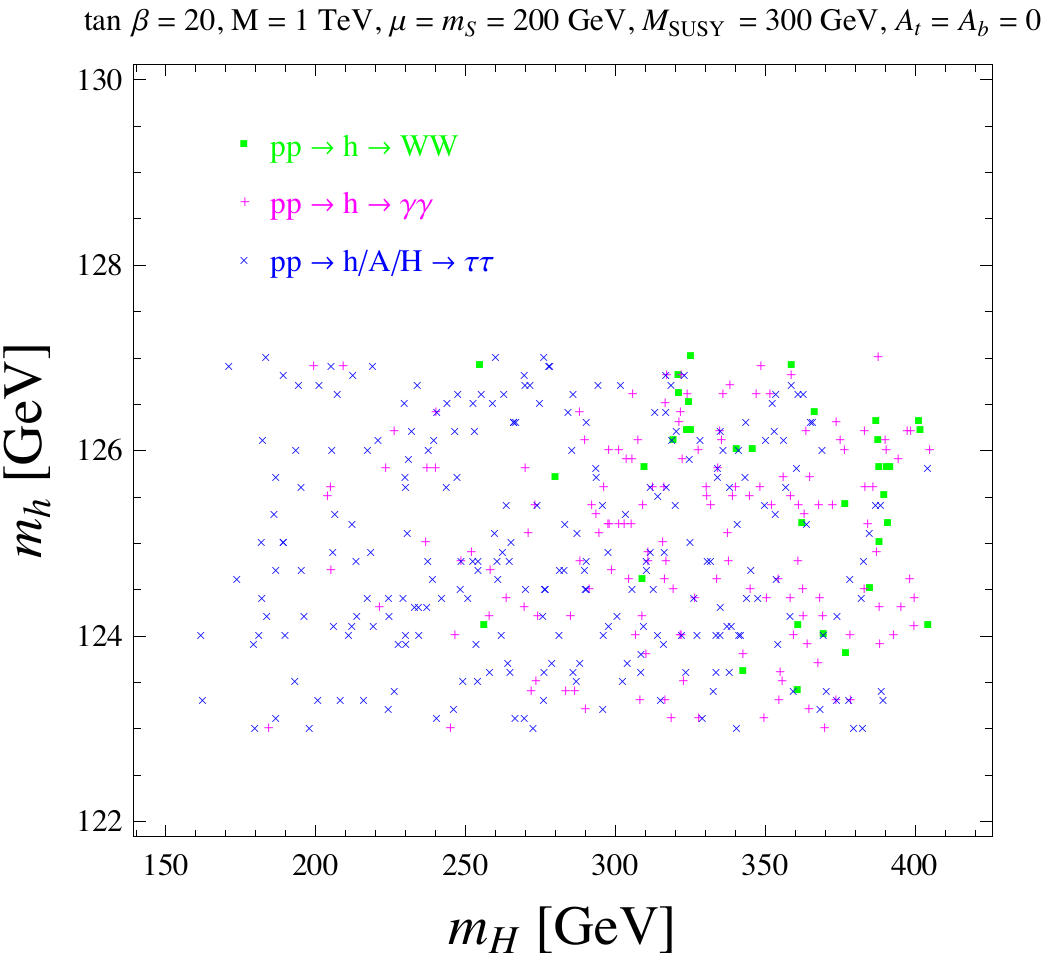,height=2.6in}
\epsfig{file=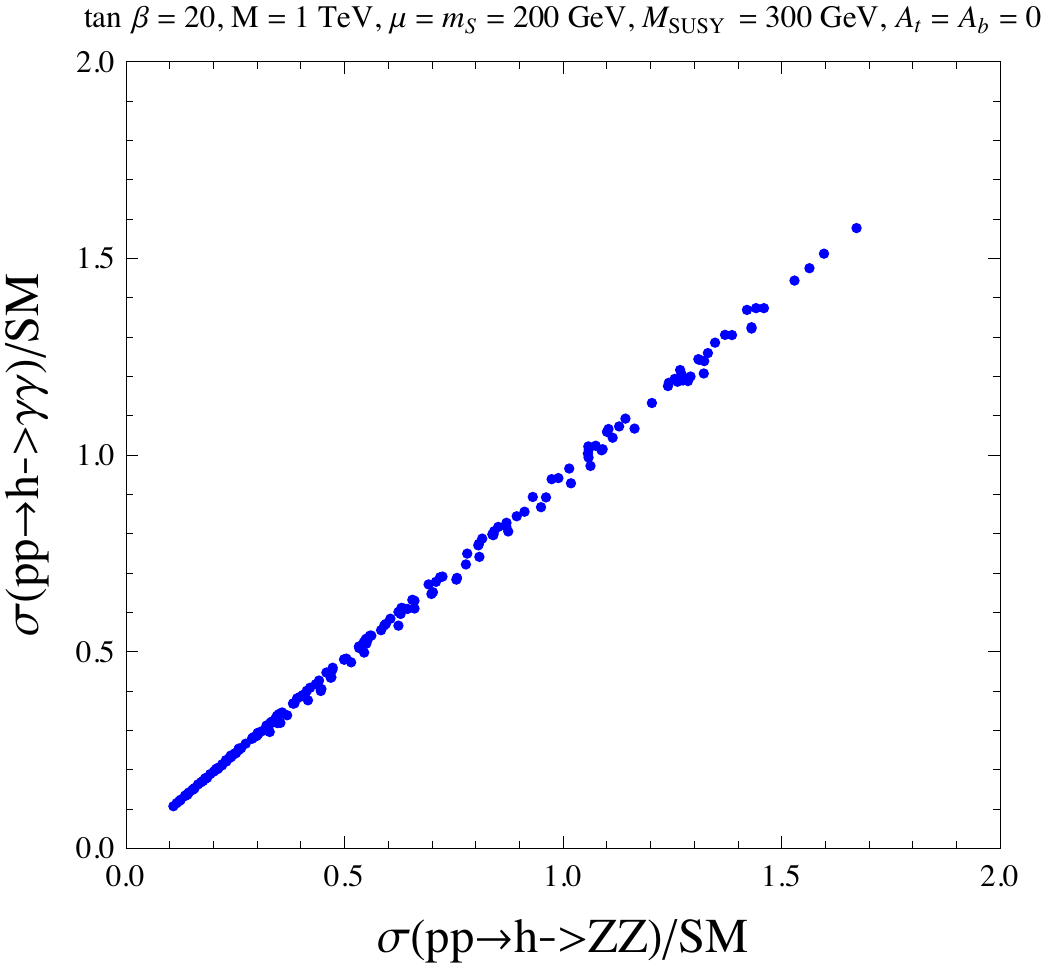,height=2.6in}
\caption{\small{a) Experimentally allowed points in the $m_h-m_H$ plane, requiring $m_h \in$ [123-127] GeV and (b) Rate to $\gamma \gamma$ vs Rate to $ZZ$ for $\tan \beta=20.$}}
\label{fig:tb20}
\end{center}
\end{figure}

In Fig.~\ref{fig:tb2} we show our scan restricted to $m_h \in[123-127]$ GeV, for $\tan \beta=2$. In the left panel we show the $m_h-m_H$ plane, while in the right panel we present the rates (cross section times branching ratio) into $\gamma\gamma$ and $ZZ$.  From the left panel we see that a Higgs boson $h$ being tested in the diphoton channel could be accompanied by a 150 GeV $H$ boson, and thus there is no need for large $m_A$. Of course this $H$ boson would be mostly gaugephobic, and thus it might be very hard to test it at the LHC. From the right panel we see that the $ZZ$ and $\gamma \gamma$ branching fractions are highly correlated, and that we can not achieve a rate to photons larger than in the SM case. 

In Fig.~\ref{fig:tb20} we show the same plot as in Fig.~\ref{fig:tb2}, but for $\tan \beta=20$. We see that now points where $h$ is tested by the diphoton channel require $m_H > 200$ GeV, while for the MSSM  one would have $m_H > 300$ GeV. In this case one can obtain rates into the diphoton channel that are larger than in the SM. The two channels are, again, highly correlated since the main contribution to $h \to \gamma \gamma$ in the SM comes from the W loop. 

In Ref.~\cite{Boudjema:2012cq} a 125 GeV Higgs in the BMSSM framework was studied with a larger detail, scanning over $\tan \beta$ and also including the large stop mixing case. It was found that the $\gamma \gamma$ and $ZZ$ channel are correlated, but the large stop mixing case changes the slope between the two rates. Hence a rate to $ZZ$ consistent with the SM allows for an enhancement of the $\gamma \gamma$ rate of about 1.5 times the SM one.

To sum up we have seen that while the MSSM can accommodate a 125 GeV Higgs boson, this requires large stop mixing, large values of $m_A$ and large $\tan \beta$. The BMSSM allows for more freedom in the supersymmetric parameter space, and in particular opens up the low $\tan \beta$ region. We have also seen that the SUSY soft parameters are crucial to disentangle the $ZZ$ and $\gamma \gamma$ channels, if this excess becomes significant in the future. 
\bigskip

I am grateful to PLHC 2012 organizers for the invitation to present this material, and for a very nice atmosphere during the conference.






 

\begin{thebibliography}{99}


\bibitem{susyexp}
https://twiki.cern.ch/twiki/bin/view/AtlasPublic/SupersymmetryPublicResults, https://twiki.cern.ch/twiki/bin/view/CMSPublic/PhysicsResultsSUS .

\bibitem{stops}
  D.~S.~M.~Alves, M.~R.~Buckley, P.~J.~Fox, J.~D.~Lykken and C.~-T.~Yu,
  arXiv:1205.5805 [hep-ph].
  Z.~Han, A.~Katz, D.~Krohn and M.~Reece,
  JHEP {\bf 1208}, 083 (2012)
  [arXiv:1205.5808 [hep-ph]].
  D.~E.~Kaplan, K.~Rehermann and D.~Stolarski,
  JHEP {\bf 1207}, 119 (2012)
  [arXiv:1205.5816 [hep-ph]].

\bibitem{stopsexp}
CMS-PAS-SUS-12-023, 
  G.~Aad {\it et al.}  [ATLAS Collaboration],
  arXiv:1208.1447 [hep-ex].
  G.~Aad {\it et al.}  [ATLAS Collaboration],
  arXiv:1208.2590 [hep-ex].
  
\bibitem{Lodone:2012kp} 
  P.~Lodone,
  Int.\ J.\ Mod.\ Phys.\ A {\bf 27}, 1230010 (2012)
  [arXiv:1203.6227 [hep-ph]].
   
   \bibitem{NaturalSUSY}
  C.~Brust, A.~Katz, S.~Lawrence and R.~Sundrum,
  JHEP {\bf 1203}, 103 (2012)
  [arXiv:1110.6670 [hep-ph]].
  M.~Papucci, J.~T.~Ruderman and A.~Weiler,
  JHEP {\bf 1209}, 035 (2012)
  [arXiv:1110.6926 [hep-ph]].
  
\bibitem{Randall:2012dm} 
  L.~Randall and M.~Reece,
  arXiv:1206.6540 [hep-ph].
  
  \bibitem{LHCHiggs}
  The ATLAS Collaboration, ATLAS Public Note ATLAS-CONF-2012-093;
  The CMS Collaboration, CMS Physics Analysis Summary CMS PAS HIG-12-020.

\bibitem{125susyhiggs}
  A.~Arbey, M.~Battaglia, A.~Djouadi, F.~Mahmoudi and J.~Quevillon,
  Phys.\ Lett.\ B {\bf 708}, 162 (2012)
  [arXiv:1112.3028 [hep-ph]].
  L.~J.~Hall, D.~Pinner and J.~T.~Ruderman,
  JHEP {\bf 1204}, 131 (2012)
  [arXiv:1112.2703 [hep-ph]].
  J.~L.~Feng, K.~T.~Matchev and D.~Sanford,
  Phys.\ Rev.\ D {\bf 85}, 075007 (2012)
  [arXiv:1112.3021 [hep-ph]].
  P.~Draper, P.~Meade, M.~Reece and D.~Shih,
  Phys.\ Rev.\ D {\bf 85}, 095007 (2012)
  [arXiv:1112.3068 [hep-ph]].
  H.~Baer, V.~Barger and A.~Mustafayev,
  Phys.\ Rev.\ D {\bf 85}, 075010 (2012)
  [arXiv:1112.3017 [hep-ph]].
  J.~-J.~Cao, Z.~-X.~Heng, J.~M.~Yang, Y.~-M.~Zhang and J.~-Y.~Zhu,
  JHEP {\bf 1203}, 086 (2012)
  [arXiv:1202.5821 [hep-ph]],
  among many others.


\bibitem{staus}
  M.~Carena, S.~Gori, N.~R.~Shah and C.~E.~M.~Wagner,
  JHEP {\bf 1203}, 014 (2012)
  [arXiv:1112.3336 [hep-ph]].
  M.~Carena, S.~Gori, N.~R.~Shah, C.~E.~M.~Wagner and L.~-T.~Wang,
  JHEP {\bf 1207}, 175 (2012)
  [arXiv:1205.5842 [hep-ph]].
  
  
\bibitem{Schael:2006cr}
  S.~Schael {\it et al.}  [ALEPH Collaboration],
  Eur.\ Phys.\ J.\  C {\bf 47}, 547 (2006)
  [arXiv:hep-ex/0602042].

  
  \bibitem{hmass}
  Y.~Okada, M.~Yamaguchi and T.~Yanagida,
  Prog.\ Theor.\ Phys.\  {\bf 85}, 1 (1991),
  J.~R.~Ellis, G.~Ridolfi and F.~Zwirner,
  Phys.\ Lett.\  B {\bf 257}, 83 (1991),
  S.~P.~Li and M.~Sher,
  Phys.\ Lett.\  B {\bf 140}, 339 (1984),
  R.~Barbieri and M.~Frigeni,
  Phys.\ Lett.\  B {\bf 258}, 395 (1991),
  M.~Drees and M.~M.~Nojiri,
  Phys.\ Rev.\  D {\bf 45}, 2482 (1992),
  J.~A.~Casas, J.~R.~Espinosa, M.~Quir\'os and A.~Riotto,
  Nucl.\ Phys.\  B {\bf 436}, 3 (1995)
  [Erratum-ibid.\  B {\bf 439}, 466 (1995)]
  [arXiv:hep-ph/9407389],
  J.~R.~Ellis, G.~Ridolfi and F.~Zwirner,
  Phys.\ Lett.\  B {\bf 262} (1991) 477,
  A.~Brignole, J.~R.~Ellis, G.~Ridolfi and F.~Zwirner,
          Phys.\ Lett.\  B {\bf 271}, 123 (1991), 
R.~Barbieri, M.~Frigeni and F.~Caravaglios,
      Phys.\ Lett.\ B {\bf 258}, 167 (1991),
   H.~E.~Haber and R.~Hempfling,
  Phys.\ Rev.\ Lett.\  {\bf 66}, 1815 (1991), 
  M.~S.~Carena, M.~Quir\'os and C.~E.~M.~Wagner,
  Nucl.\ Phys.\  B {\bf 461}, 407 (1996)
  [arXiv:hep-ph/9508343].
      
 \bibitem{dim5}     
  A.~Brignole, J.~A.~Casas, J.~R.~Espinosa and I.~Navarro,
  Nucl.\ Phys.\  B {\bf 666}, 105 (2003)
  [arXiv:hep-ph/0301121].
  M.~Dine, N.~Seiberg and S.~Thomas,
  Phys.\ Rev.\  D {\bf 76}, 095004 (2007)
  [arXiv:0707.0005 [hep-ph]].
  
  \bibitem{hdo}
  A.~Strumia,
  Phys.\ Lett.\  B {\bf 466}, 107 (1999)
  [arXiv:hep-ph/9906266],
  I.~Antoniadis, E.~Dudas and D.~M.~Ghilencea,
  JHEP {\bf 0803}, 045 (2008)
  [arXiv:0708.0383 [hep-th]],
  I.~Antoniadis, E.~Dudas, D.~M.~Ghilencea and P.~Tziveloglou,
  Nucl.\ Phys.\  B {\bf 808}, 155 (2009)
  [arXiv:0806.3778 [hep-ph]],
  P.~Batra and E.~Pont\'on,
  Phys.\ Rev.\  D {\bf 79}, 035001 (2009)
  [arXiv:0809.3453 [hep-ph]],
  K.~Blum, C.~Delaunay and Y.~Hochberg,
  Phys.\ Rev.\  D {\bf 80}, 075004 (2009)
  [arXiv:0905.1701 [hep-ph]],
  J.~A.~Casas, J.~R.~Espinosa and I.~Hidalgo,
  JHEP {\bf 0401}, 008 (2004)
  [arXiv:hep-ph/0310137],
  S.~Cassel, D.~M.~Ghilencea and G.~G.~Ross,
  Nucl.\ Phys.\  B {\bf 825}, 203 (2010)
  [arXiv:0903.1115 [hep-ph]],

  
\bibitem{Carena:2009gx}
  M.~Carena, K.~Kong, E.~Pont\'on and J.~Zurita,
  Phys.\ Rev.\  D {\bf 81}, 015001 (2010)
  [arXiv:0909.5434 [hep-ph]].
  
\bibitem{Antoniadis:2009rn}
  I.~Antoniadis, E.~Dudas, D.~M.~Ghilencea and P.~Tziveloglou,
  Nucl.\ Phys.\  B {\bf 831}, 133 (2010)
  [arXiv:0910.1100 [hep-ph]],

\bibitem{CPZ}  
  M.~Carena, E.~Ponton and J.~Zurita,
  Phys.\ Rev.\ D {\bf 82}, 055025 (2010)
  [arXiv:1005.4887 [hep-ph]].
  M.~Carena, E.~Ponton and J.~Zurita,
  Phys.\ Rev.\ D {\bf 85}, 035007 (2012)
  [arXiv:1111.2049 [hep-ph]].



  \bibitem{Boudjema:2011aa} 
  F.~Boudjema and G.~Drieu La Rochelle,
  Phys.\ Rev.\ D {\bf 85}, 035011 (2012)
  [arXiv:1112.1434 [hep-ph]].
\bibitem{Boudjema:2012cq} 
  F.~Boudjema and G.~D.~La Rochelle,
  Phys.\ Rev.\ D {\bf 86}, 015018 (2012)
  [arXiv:1203.3141 [hep-ph]].
  
 
  

  
\end{thebibliography}
\end{document}